\documentclass[aps,prd,twocolumn,preprintnumbers,superscriptaddress,floatfix]{revtex4}

\usepackage{multirow, graphicx,amssymb,url,mathrsfs,amsmath}
\usepackage{eucal,wrapfig,boxedminipage,setspace,subfigure}
\usepackage{amsxtra,amstext,latexsym,dsfont}

\def\IR{{\hbox{{\rm I}\kern-.2em\hbox{\rm R}}}}
\def\IB{{\hbox{{\rm I}\kern-.2em\hbox{\rm B}}}}
\def\IN{{\hbox{{\rm I}\kern-.2em\hbox{\rm N}}}}
\def\IC{\,\,{\hbox{{\rm I}\kern-.59em\hbox{\bf C}}}}
\def\IZ{{\hbox{{\rm Z}\kern-.4em\hbox{\rm Z}}}}
\def\IP{{\hbox{{\rm I}\kern-.2em\hbox{\rm P}}}}
\def\IH{{\hbox{{\rm I}\kern-.4em\hbox{\rm H}}}}
\def\ID{{\hbox{{\rm I}\kern-.2em\hbox{\rm D}}}}

\newcommand{\beq}{\begin{equation}}
\newcommand{\eeq}{\end{equation}}
\newcommand{\bea}{\begin{eqnarray}}
\newcommand{\eea}{\end{eqnarray}}

\begin{document}

\voffset 1cm

\newcommand\sect[1]{\emph{#1}---}

\title{A Holographic Study of the Gauged NJL Model}

\author{Will Clemens, Nick Evans}
\affiliation{ STAG Research Centre \&  Physics and Astronomy, University of
Southampton, Southampton, SO17 1BJ, UK}

\begin{abstract}

\noindent The Nambu Jona-Lasinio model of chiral symmetry breaking predicts a second order chiral phase transition. 
If the fermions in addition have non-abelian gauge interactions then the transition is expected to become a crossover 
as the NJL term enhances the IR chiral symmetry breaking of the gauge theory. We study this behaviour in the holographic
Dynamic AdS/QCD description of a non-abelian gauge theory with the NJL interaction included using Witten's multi-trace
prescription. We study the behaviour of the mesonic spectrum as a function of the NJL coupling and the ratio of the UV 
cut off scale to the dynamical scale of the gauge theory. 

\end{abstract}

\maketitle

\newpage

The gauged Nambu Jona-Lasinio  (NJL) model \cite{Yamawaki:1996vr}
\begin{equation} {\cal L} = {1 \over 4 g^2} F^{\mu \nu} F_{\mu \nu} + i \bar{q} D \hspace{-0.25cm} \slash  ~~ q  + {g^2 \over \Lambda^2} (\bar{q}_L q_R \bar{q}_R q_L + h.c.), \end{equation}
with a UV cut off $\Lambda$, provides an interesting study of two competing chiral symmetry breaking interactions. The NJL term triggers symmetry breaking  
at a second order transition if the coupling exceeds a critical value of $g = 2 \pi$ \cite{Nambu:1961tp}. The QCD-like non-abelian gauge interactions run from asymptotic freedom to strong coupling in the infrared (IR) and again trigger chiral symmetry breaking \cite{Gross:1973ju}. In combination the two mechanisms enhance each other and the NJL transition is changed to a cross-over as the dynamics change from being dominated by the IR gauge theory to the ultraviolet (UV) NJL term \cite{Yamawaki:1996vr}. Over the years there has been interest in the model for possible roles in strongly coupled beyond the standard model physics but it remains an interesting system in itself. It has been studied previously using gap equation \cite{Appelquist:1988as,Miransky:1989nu,Takeuchi:1989qa,King:1990hd} and lattice techniques \cite{Catterall:2011ab,Rantaharju:2016jxy}.

Here we wish to study the model holographically \cite{Maldacena:1997re}. Quarks with couplings to conformal ${\cal N}=4$ gauge dynamics can be described, in the quenched limit, by probe D7 branes in AdS$_5 \times S^5$ \cite{Karch:2002sh}. The Dirac Born Infeld (DBI) action provides a very simple description of the mesonic spectrum. In a very small number of cases supersymmetry can be broken by, for example, a background U(1)$_B$ magnetic field \cite{Filev:2007gb} or a running dilaton \cite{Babington:2003vm} and chiral symmetry breaking is described. The mechanism is simplest to understand if one expands around the chirally symmetric vacuum \cite{Alvares:2012kr} (see also the work in \cite{Jarvinen:2011qe,Kutasov:2011fr,Jarvinen:2009fe}). The action becomes that of a scalar in AdS (the brane embedding) with a mass squared that runs with radial distance in AdS (ie with energy scale) and the instability to chiral symmetry breaking sets in when the Breitenlohner Freedman (BF) bound \cite{Breitenlohner:1982jf} is violated in the IR. Using the AdS/CFT dictionnary \cite{Maldacena:1997re} this translates to saying that the anomalous dimension of the quark condensate, $\gamma$, runs into the IR until $\gamma=1$, at which point chiral symmetry breaking occurs. 

It is irrisitible to take this very simple model of the quark dynamics and replace the running $\gamma$ with a sensible guess for another theory where a full holographic description of the glue background does not exist (note such a background might include backreaction of the quarks themselves). The Dynamic AdS/QCD model \cite{Alho:2013dka,Evans:2013vca} is such a  model with the perturbative running for an SU($N_c$) gauge theory with $N_f$ quarks inserted. If one simply takes the one-loop expressions for $\gamma$ for $N_c=3$ and $N_f=2$ naively extended to the scale where $\gamma=1$, a very sensible description of QCD is obtained (inspite of the very simplified choice of operators involved in the dynamics and the neglect of any stringy physics of the dual). This is the holographic description of the gauge theory we will use here.

The NJL interaction may be included using Witten's double trace prescription \cite{Witten:2001ua} and it has recently been shown how the basic NJL second order transition can be achieved holographically in non-supersymmetric duals \cite{Evans:2016yas}. Thus we will be able to display here the behaviour in a number of mesonic variables of the parameter space of the gauged NJL model. Below we will introduce Dynamic AdS/QCD, Witten's prescription and then present our computations for the model. \vspace{-0.6cm}

\section{Dynamic AdS/QCD}
\vspace{-0.5cm}

Quarks can be associated in AdS/CFT with strings stretching between a probe D7 brane and the D3 branes from which the ${\cal N}=4$ SYM gauge fields originate \cite{Karch:2002sh}. In the probe approximation the D7 branes lie in the $x_{3+1}$ field theory directions, $\rho$ and $\Omega_3$ coordinates of the transverse space in the AdS$_5 \times S^5$ metric
\begin{equation} ds^2 = r^2 dx_{3+1}^2 + {1 \over r^2} ( d\rho^2 + \rho^2 d \Omega_3^2 + dX_{89}^2 ), ~~~~~~~r^2 = \rho^2 + |X|^2 \end{equation}  
A flat D7 brane that intersects the D3 at the origin would lie at $|X|=0$ and describe massless quarks. The DBI action is 
\begin{equation} S_{D7}  = - T_7 \int d^8\xi e^{\phi} \sqrt{-{\rm det}G_{ab} + F_{ab}} \end{equation}
where the dilaton factor is constant in pure AdS but  is non-zero in the presence, for example, of a supersymmetry breaking U(1)$_B$ magnetic field \cite{Filev:2007gb}  or one could imagine it as some running gauge coupling \cite{Babington:2003vm}. One might now substitute the metric, integrate over the inert $\Omega_3$ and expand in small $X$ to study the instability of the brane from the flat massless state \cite{Alvares:2012kr}. One essentially finds \cite{Alho:2013dka,Evans:2013vca} \vspace{-0.5cm}

\begin{figure}[]
\centering
\includegraphics[width=6cm]{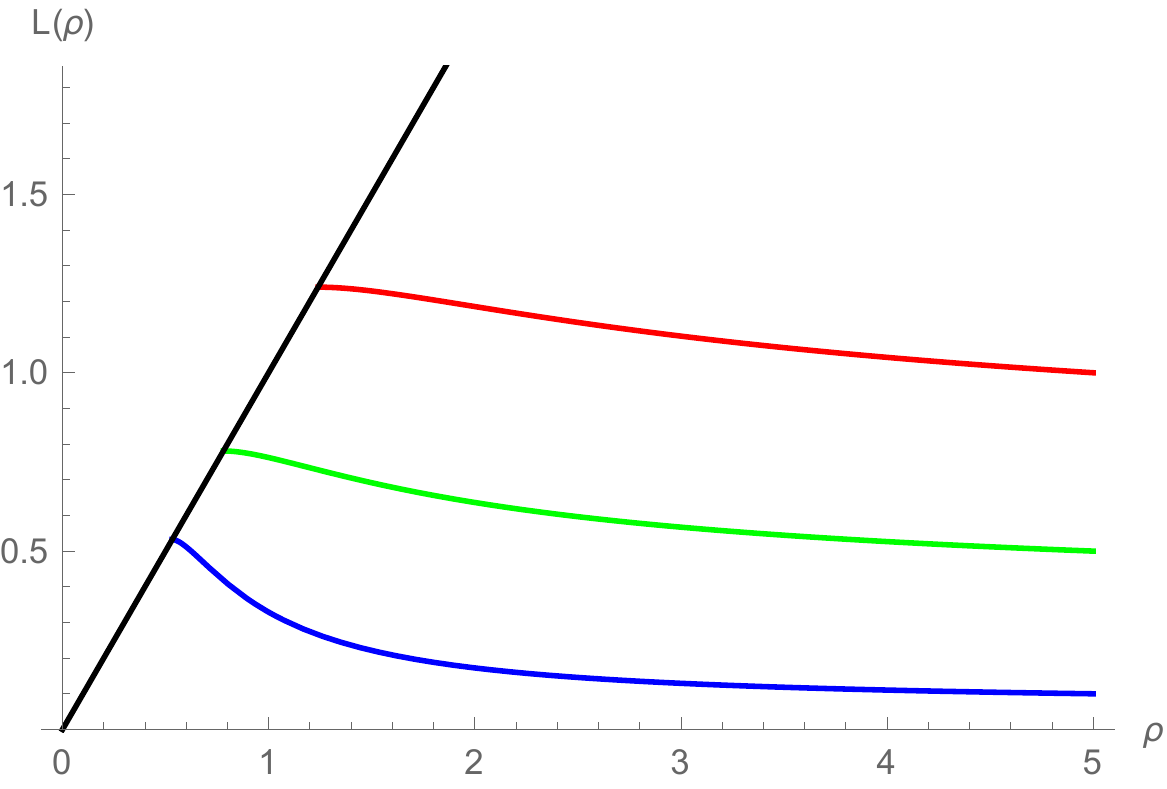}
\caption{The embeddings for different IR choice $L_0$. Here the scale is set by the value of $r= \Lambda_{BF}$ at which the BF bound is violated (an approximation to $\Lambda_{QCD}$ being set to 1).  }
\label{embedding}
\end{figure}

\begin{equation} \begin{array}{lcl}
S & = & -\int d^4x~ d \rho\, {\rm{Tr}}\, \rho^3 
\left[  {1 \over r^2} |D X|^2 \right.  \\ &&\\
&& \left. \hspace{1cm}  +  {\Delta m^2(r) \over \rho^2} |X|^2   + {1 \over 2 \kappa^2} (F_V^2 + F_A^2) \right], 
\label{daq} \end{array}
\end{equation}
$\Delta m^2$ is just the expansion coefficient from the dilaton term.  There are some sleights of hand here to generate a consistent phenomenological model: \vspace{-0.4cm}

\begin{itemize}
\item One should, formally, expand the factors of $r$ (which contain $X$) but they are retained for two reasons. If we simply write $\Delta m^2(\rho)$ then if this term trips us through the BF bound in some range of small $\rho$ then the instability would exist however large $|X|=L$ were to grow. If we instead promote $\rho$ in this term to $r$ then at sufficiently large $L$ the instability is no longer present and a full solution for $L$ is possible. In addition when one looks at fluctuations about the background embedding those states only know about the non-trivial embedding through the factors of $L$ in the metric - these terms communicate the existence of a gap to the bound states.  We are of course being inconsistent about higher order terms in $X$ but the key point is to keep the $X^2$ term that triggers the BF bound violation and the brutality of our other assumptions will be judged by the success of the model below.
\item We have added an axial gauge field in symmetry with the vector field whilst in the full D3/D7 system this state is more complex. 
\item We have allowed a free 5d gauge coupling $\kappa$ which we expect to become a phenomenological parameter with the breaking of supersymmetry. 
\end{itemize} \vspace{-0.4cm}

We will impose appropriate gauge dynamics on our model by fixing the form of $\Delta m^2(r)$ using the one loop QCD runing with $N_c=3$ and $N_f=2$.
\begin{equation} 
\mu { d \alpha \over d \mu} = - b_0 \alpha^2, \hspace{1cm}
 b_0 = {1 \over 6 \pi} (11 N_c - 2N_F), \end{equation}
The one loop result for the anomalous dimension is
\begin{equation} \gamma = {3 C_2 \over 2\pi}\alpha= {3 (N_c^2-1) \over 4 N_c \pi} \alpha\,.  \end{equation}

We will identify the RG scale $\mu$ with the AdS radial parameter $r$ in our model. 
Working perturbatively from the AdS result $m^2 = \Delta(\Delta-4)$ \cite{Maldacena:1997re} we have
\begin{equation} \label{dmsq3} \Delta m^2 = - 2 \gamma = -{3 (N_c^2-1) \over 2 N_c \pi} \alpha\, .\end{equation}
We call the scale where the BF bound is first violated $\Lambda_{BF}$ and it is a rough measure of the traditional QCD scale $\Lambda_{QCD}$

The vacuum structure of the theory is found by setting all fields except $|X|=L$ to zero. The Euler-Lagrange equation for the determination of $L$, in the case of a constant $\Delta m^2$, is 
\begin{equation} \label{embedeqn}
\partial_\rho[ \rho^3 \partial_\rho L]  - \rho \Delta m^2 L  = 0\,. \end{equation}
We can now ansatz the $r$ dependent $\Delta m^2$ above to describe the running of the dimension of $\bar{q}q$ (we do this at the level of the equation of motion).
To find numerical solutions we need an IR boundary condition. In top down models $L'(0)=0$ is the condition for a regular solution.  Since we do not wish to describe IR physics below the quark mass (where the quark contribution to the running coupling will decouple) we use a very similar on-shell condition - we shoot from points $L(\rho=L_0) = L_0$ with $L'(L_0)=0$. In the UV the solution (neglecting $\Delta m^2$ which falls close to zero) takes the form 
\begin{equation} L = m + {c \over \rho^2} \label{UVform} \end{equation} where $m$ in interpreted as the UV quark mass and $c$ as the quark condensate. We show sample embeddings in Fig \ref{embedding} for different choices of $L_0$.

The spectrum of the theory is determined by looking at linearized fluctuations of the fields about the vacuum where fields generically take the 
form $f(\rho)e^{ip.x}, p^2=-M^2$. A Sturm-Louville equation results for $f(\rho)$ leading to a discrete spectrum. By substituting the wave functions back into the action and integrating over $\rho$ the decay constants can also be determined.
The normalizations of the fluctuations are determined by matching to the gauge theory expectations for the VV, AA and SS correlators in the UV of the theory. This full procedure is described in detail in \cite{Alho:2013dka}.

With $N_c$ and $N_f$ fixed the free parameters in the theory are the overall scale $\Lambda_{BF}$, the UV quark mass and the 5d coupling $\kappa$. For example one can fix $\Lambda_{BF}$ by scaling to give the correct $m_\rho$; the remaining parameters can then be fitted to the data. We choose to minimize the maximum deviation $|\delta O|/ O$ in any observable and find a good fit at $m_{UV}=0.05 \Lambda_{BF}$ at a UV scale of $5 \Lambda_{BF}$ and $\kappa =76$: \vspace{-0.5cm}

\begin{center}
\begin{tabular}{|c|c|c|}
\hline
 & Model & QCD \\  \hline
 $m_\rho$& 775 MeV & 775 MeV \\
 $m_{a_1}$ & 1467 MeV & 1230 MeV \\
 $m_\sigma$ & 981 MeV & 500 MeV \& 980 MeV \\
 $F_\rho$ & 311 MeV & 345 MeV \\
 $F_{a_1}$ & 390 MeV  & 433 MeV \\
 $ f_\pi$ & 77 MeV  & 92 MeV \\
\hline  
\end{tabular}
\end{center} \vspace{-0.5cm}

All the mesonic variables lie within $20\%$ of the experimental centrepoints shown except for the $\sigma$ meson mass that lies very close to the first excited $f_0(980)$ state. The lighter $f_0(500)$ is thought to potentially be a mesonic molecule \cite{Londergan:2013dza} which might explain the discrepancy. In anycase our model provides a sufficiently close spectrum match to begin an analysis of NJL dynamics in the model.  

\begin{figure}[]
\centering
\includegraphics[width=6cm]{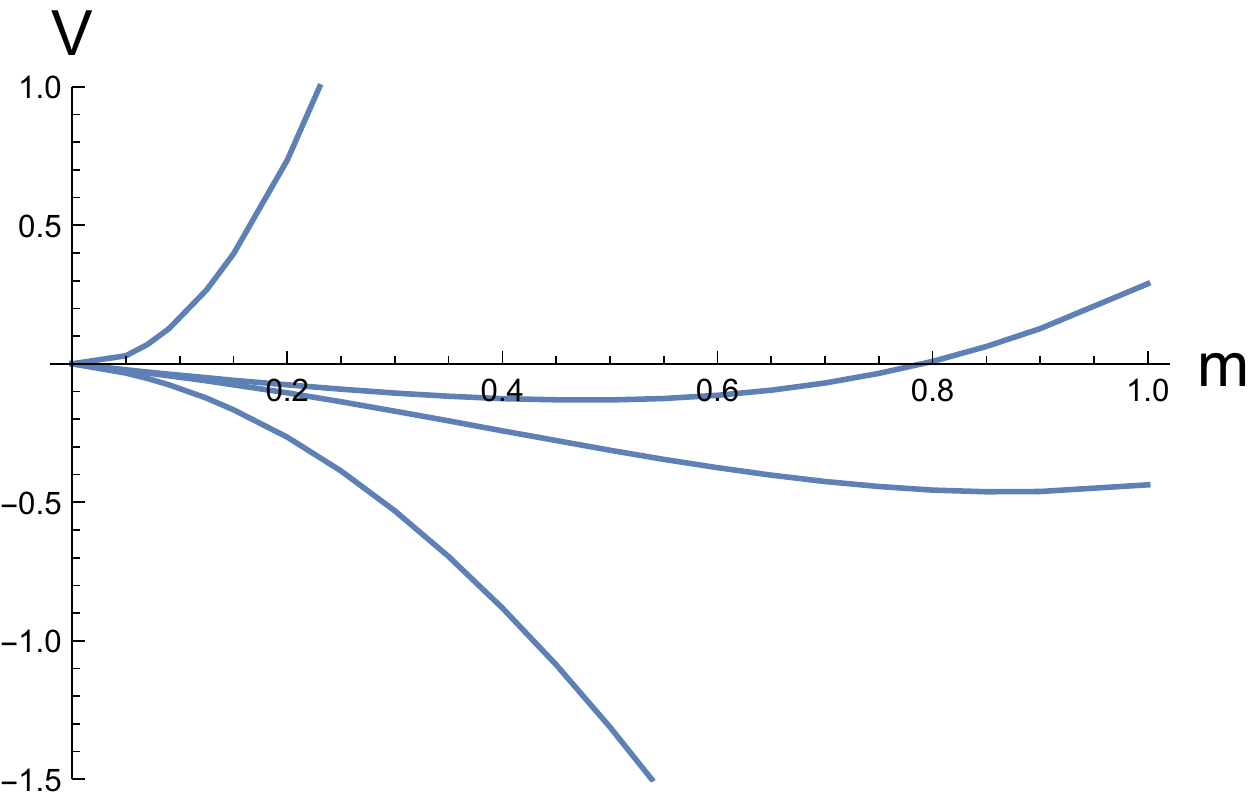}
\caption{Plots of the potential against the UV quark mass: the lower curve is that of the underlying gauge theory without an NJL term and is unbounded. Moving up we have added the term $\Lambda^2 m^2/g^2$ with $g=2.5,2.3,1$ from bottom to top. The addition of an NJL term generates a minimum of the potential that tracks to $m=0$ at $g=0$.
All dimensionful objects are expressed in terms of $\Lambda_{BF}$.  }
\label{potential}
\end{figure}

\section{NJL Interactions}
\vspace{-0.5cm}

Consider a free fermion with a four fermion interaction $g^2/\Lambda^2 \bar{q}_L q_R \bar{q}_R q_L$. In the standard NJL approximation there are two contributions to the effective potential \cite{JonaLasinio:1964cw}. First there is the one loop Coleman Weinberg potential \cite{Coleman:1973jx} for the free quarks 
\begin{equation} V_{\rm eff} = - \int^\Lambda_0 {d^4k \over (2 \pi)^4} Tr \log (k^2 +m^2) \label{ColemanW}\end{equation}
This falls with growing $m$ and is unbounded, although normally one treats $m$ as a fixed parameter so one would not seek to minimize this potential. When we add the four fermion term we allow $m$ to become dynamically determined but there is the extra term from the four fermion interaction evaluated on $m= (g^2/\Lambda^2) \langle \bar{q} q \rangle$
\begin{equation} \Delta V_{\rm eff} = {\Lambda^2 m^2 \over g^2} \label{NJLextra} \end{equation}
This makes the effective potential bounded and ensures a minimum. For small $g$ the extra term is large and the minimum is at $m=0$. When $g$ rises above $2 \pi$ the minimum lies away from $m=0$. The phase transition is second order.

In Witten's prescription for ``multi-trace'' operators \cite{Witten:2001ua} we add the equivalent of the extra potential term (\ref{NJLextra}) as a boundary term at the UV cut off $\Lambda$. For large $\Lambda$ where $L\simeq m$ the term we add is
\begin{equation} \Delta S _{UV}=   {L^2 \Lambda^2 \over g^2  } \label{bound} \,.  \end{equation}
The effective potential from the background model is computed by evaluating minus the action (\ref{daq}) evaluated on the vacuum solution as a function of the UV mass term. We extract the values of $m$ and $c$ in the UV by fitting to the form (\ref{UVform}) near the cut off. In Fig \ref{potential} we plot the effective potential in the holographic description of the $N_c=3$ and $N_f=2$ gauge theory showing that like (\ref{ColemanW}) it is unbounded with $m$. When we add the potential term in (\ref{bound}) a minimum at non-zero $m$ is found. The minimum tracks to $m=0$ at $g=0$ indicating the crossover nature of the transition.

\begin{figure}[]
\centering
\includegraphics[width=8cm]{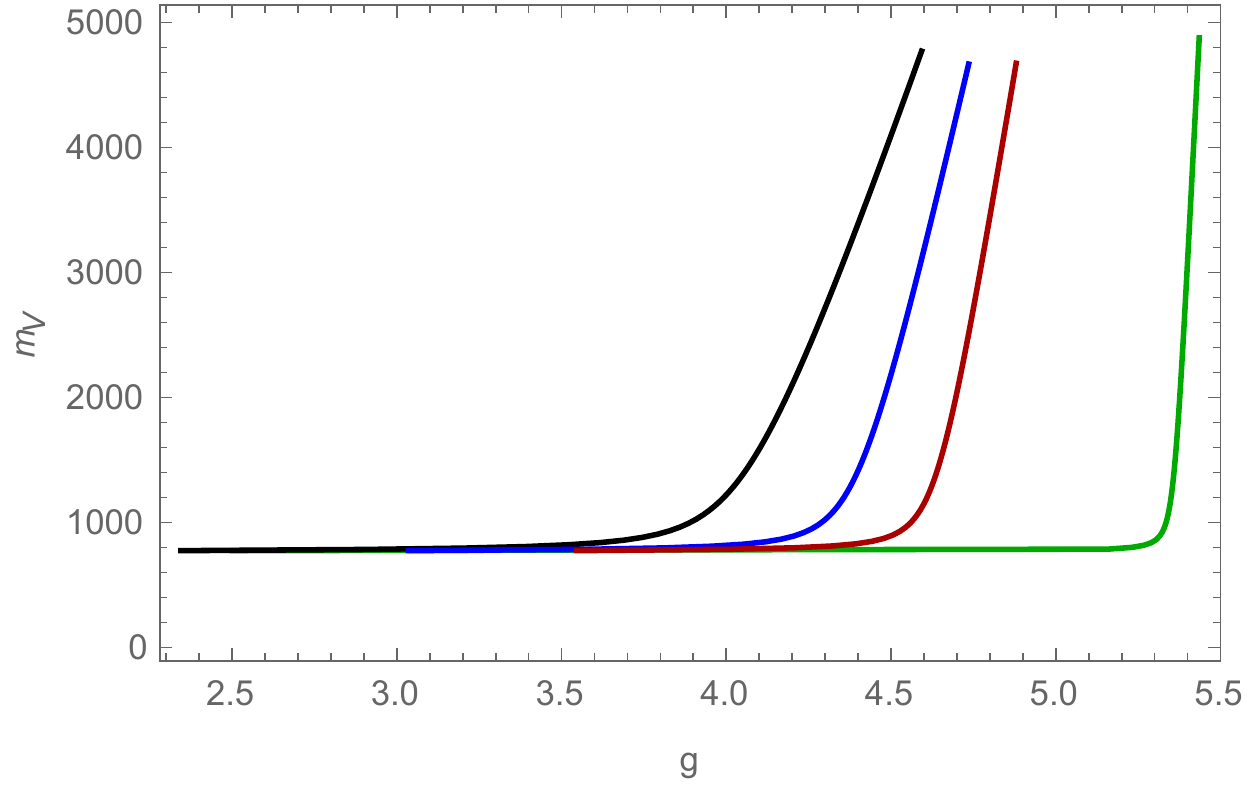}  \\  \includegraphics[width=8cm]{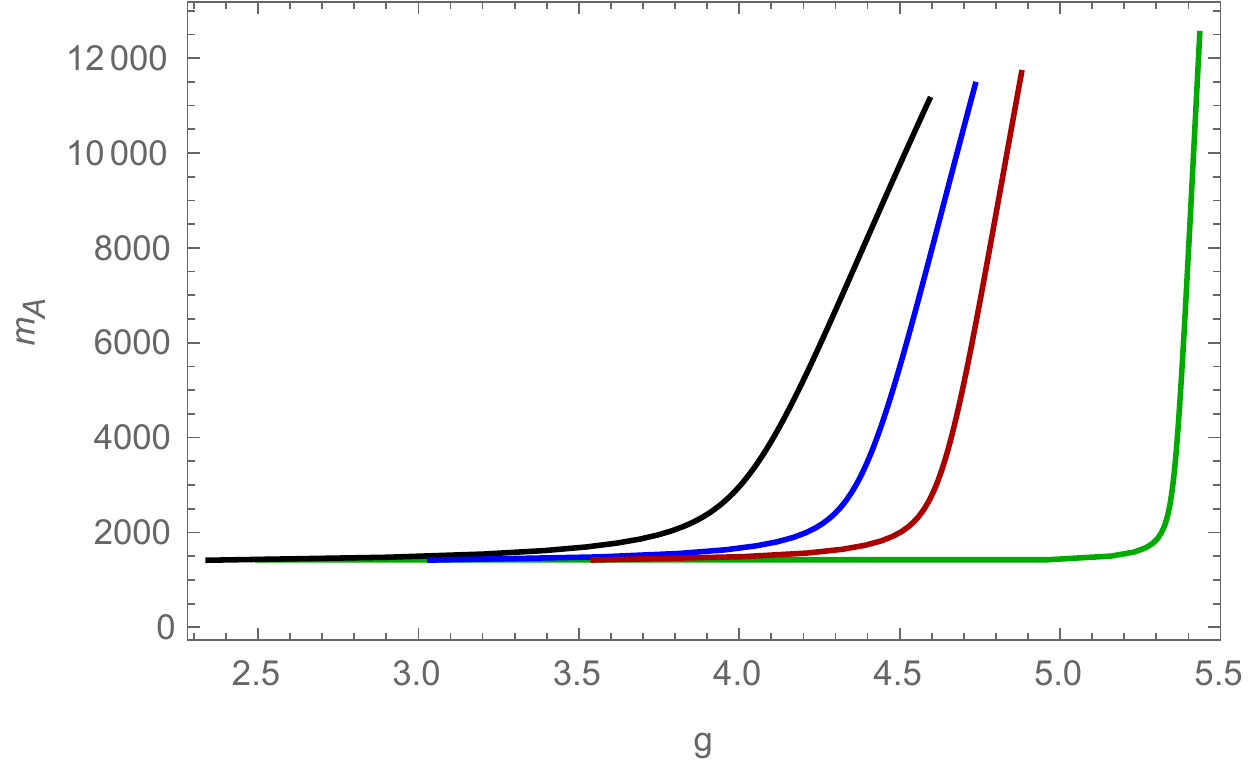}  \\  \includegraphics[width=8cm]{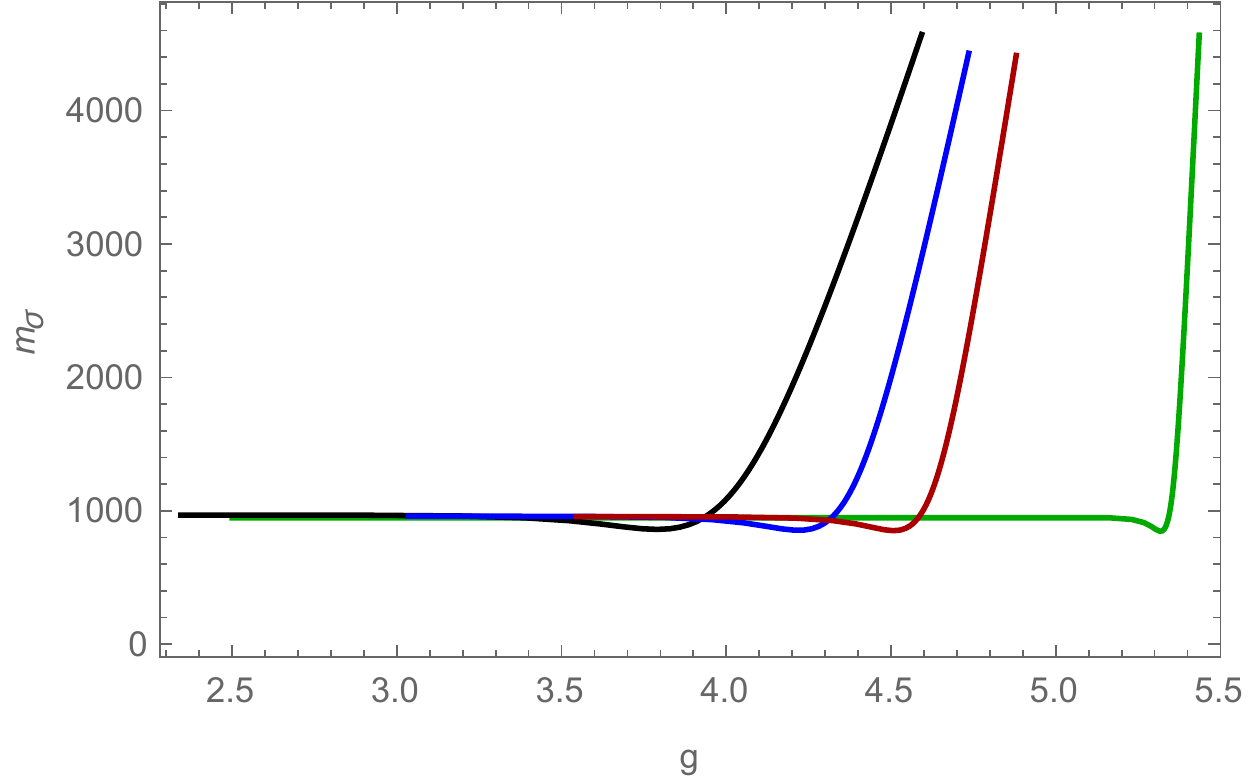}
\caption{Plots showing the vector and axial vector meson ($\rho$ and $a_1$) and $\sigma$ meson ($f_0$) masses against the NJL coupling constant $g$ for choices of UV cut off $\Lambda = 10,15,20,50 \Lambda_{BF}$. }
\label{mvvsg}
\end{figure}

We can also understand Witten's prescription in terms of a change to the UV boundary condition on the solution of the embedding equation. Varying the action gives
\begin{equation}  \delta S = 0 = -\int d \rho \left(\partial _\rho{\partial {\cal L }\over \partial L'}  - {\partial {\cal L }
\over \partial L} \right)  \delta L  + \left. {\partial {\cal L }\over \partial L'} \delta L \right|_{{UV, IR}} \,.  \end{equation}
Normally in the UV one would require the mass to be fixed and $\delta L=0$ to satisfy the boundary condition but now we allow $L$ to change and instead impose 
\begin{equation} 0 = {\partial {\cal L} \over \partial L'}  + {2 L \Lambda_{UV}^2 \over g^2 }  \,,   \end{equation}
where we have included the variation of the surface term. For our action ${\partial {\cal L} \over \partial L'} = \rho^3 \partial_\rho L$. Assuming (\ref{UVform}) we find that we need
\begin{equation} m \simeq {g^2 \over \Lambda^2} c \end{equation}
This condition is simpler to apply to the solutions of the Euler Lagrange equation than constructing and minimizing the effective potential but equivalent.

\section{Results for the Gauged NJL Model}

\begin{figure}[]
\centering
\includegraphics[width=8cm]{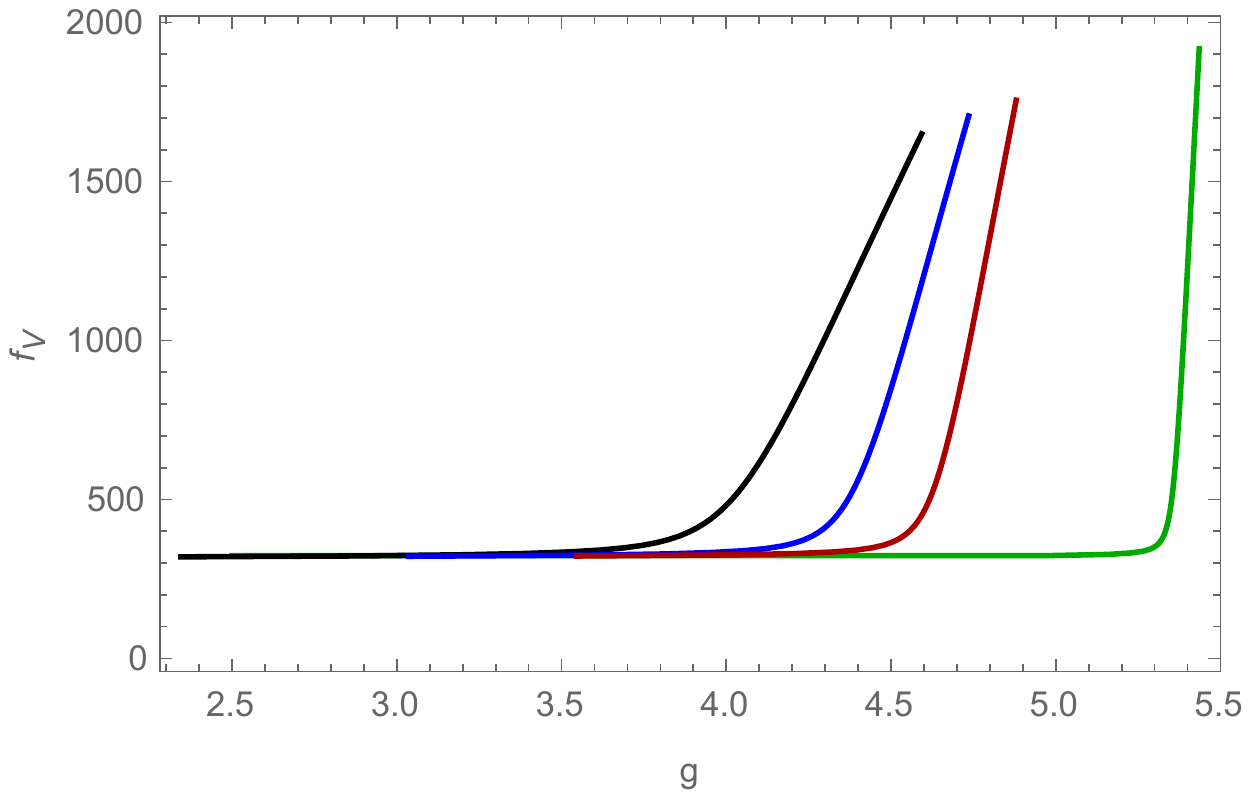}  \\  \includegraphics[width=8cm]{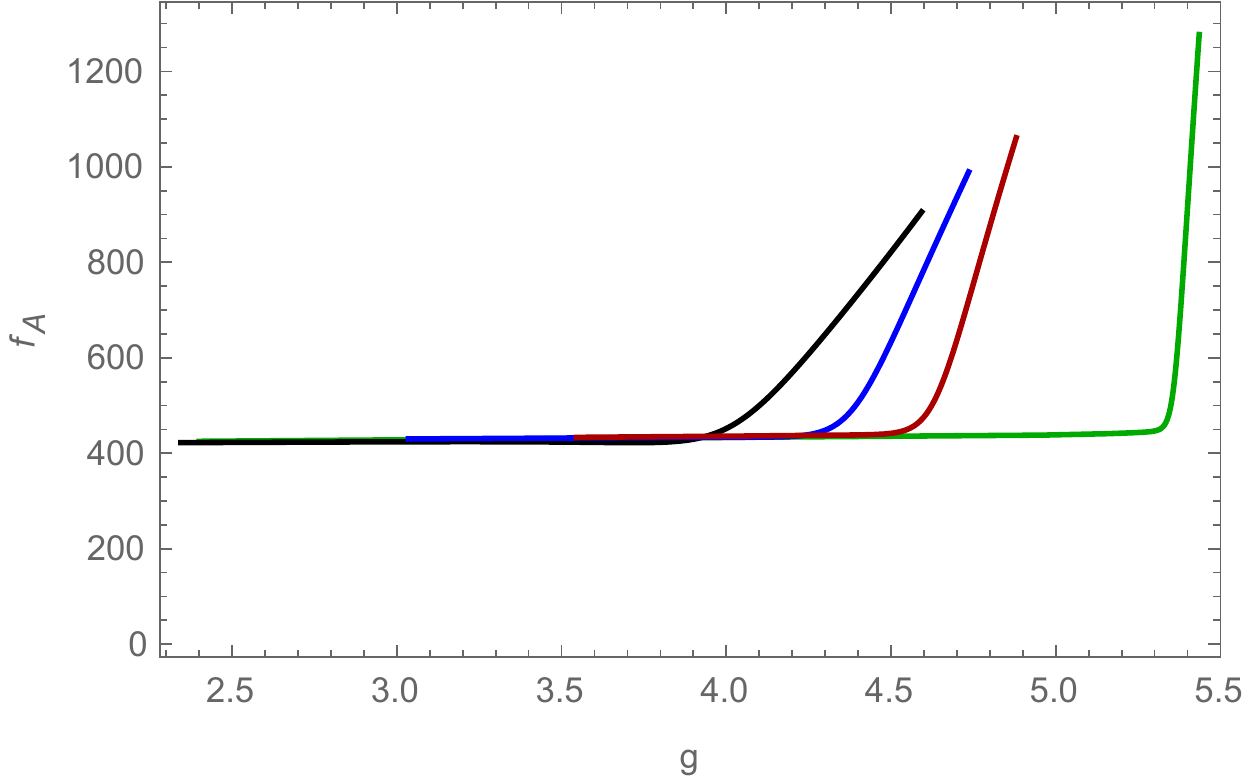} \\  \includegraphics[width=8cm]{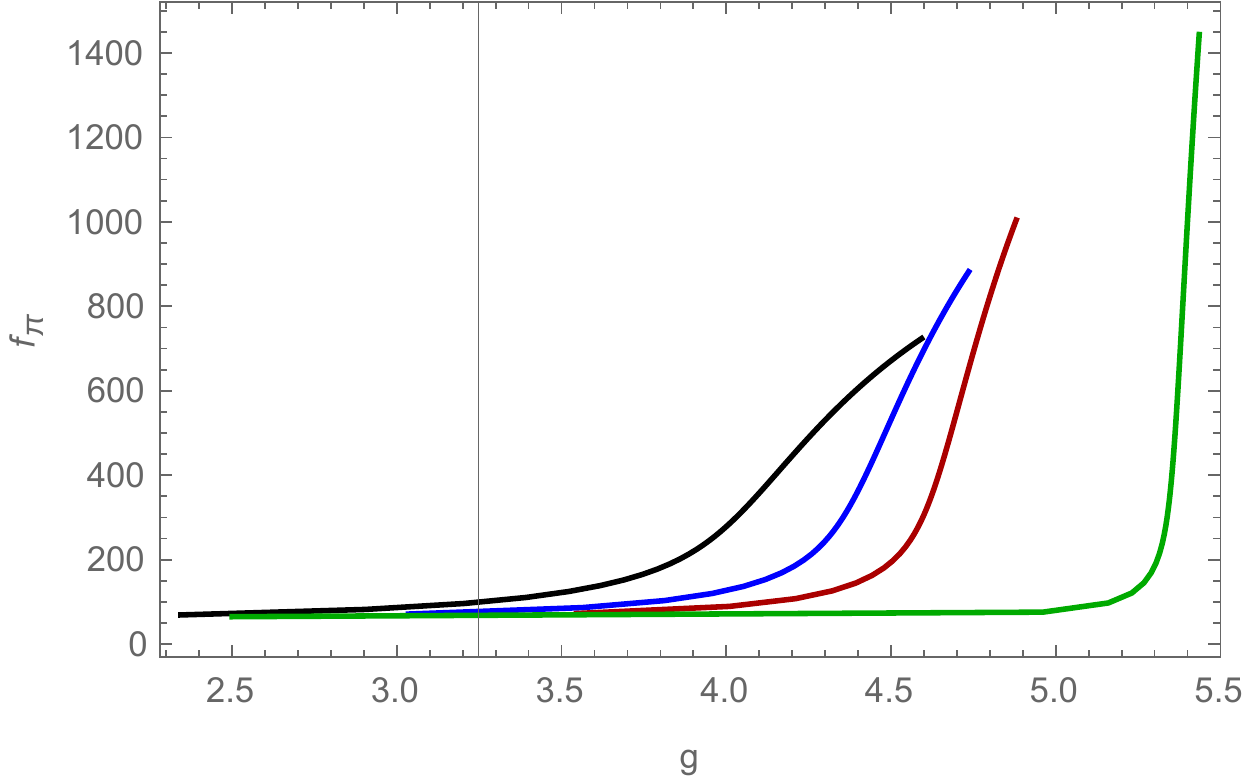}
\caption{Plot showing the vector and axial vector meson  and $\pi$ meson  decay constants against the NJL coupling constant $g$ for choices of UV cut off $\Lambda = 10,15,20,50 \Lambda_{BF}$. }
\label{fvvsg}
\end{figure}

We can now study the mesonic variables of our holographic model in the presence of an NJL interaction. In Fig \ref{mvvsg} and Fig \ref{fvvsg} we display respectively the meson masses ($m_\rho, m_{a_1}, m_\sigma$) and the decay constants ($F_\rho, F_{a_{1}}, f_\pi$) of the SU(3) gauge theory with $N_f=2$. When the NJL coupling $g=0$ the model generates the output of the table above and corresponds to our description of QCD.  Results are shown in the figures for cut offs $\Lambda = 10,15,20,50 \Lambda_{BF}$. As $g$ grows at the cut off the NJL interaction enhances the mass gap of the theory.  At  larger $g$ the NJL interaction dominates and generates a much larger gap that sharply grows to the cut off scale. This clearly shows the anticipated cross over behaviour. As the cut off is taken large relative to $\Lambda_{QCD}$ the transition becomes sharper and moves closer to the second order NJL only behaviour. The holographic model reproduces the expected physics well. Interestingly there is a small dip in the sigma meson mass before the NJL interaction dominates although it is not a large effect. In Fig \ref{lambda=20,50} we show the full set of observables, normalized byt heir vlaue at $g=0$, against $g$ for two different $\Lambda$ to stress the sharpening of the transition with growing cut off and show the relative growth of the observables.

The success in finding a holographic description of the gauged NJL model opens up the possibility of doing a range of beyond the standard model physics including extended techniolour \cite{Bardeen:1989ds}
 and top condensation \cite{Farhi:1980xs} interactions. We explore the phenomenology of these models in detail in the paper \cite{partner}. 
 \bigskip \bigskip

\noindent{\bf Acknowledgements:} NE's work was supported by the STFC consolidated grant ST/L000296/1 and WC's by an STFC studentship.

\begin{figure}[]
\centering
\includegraphics[width=8cm]{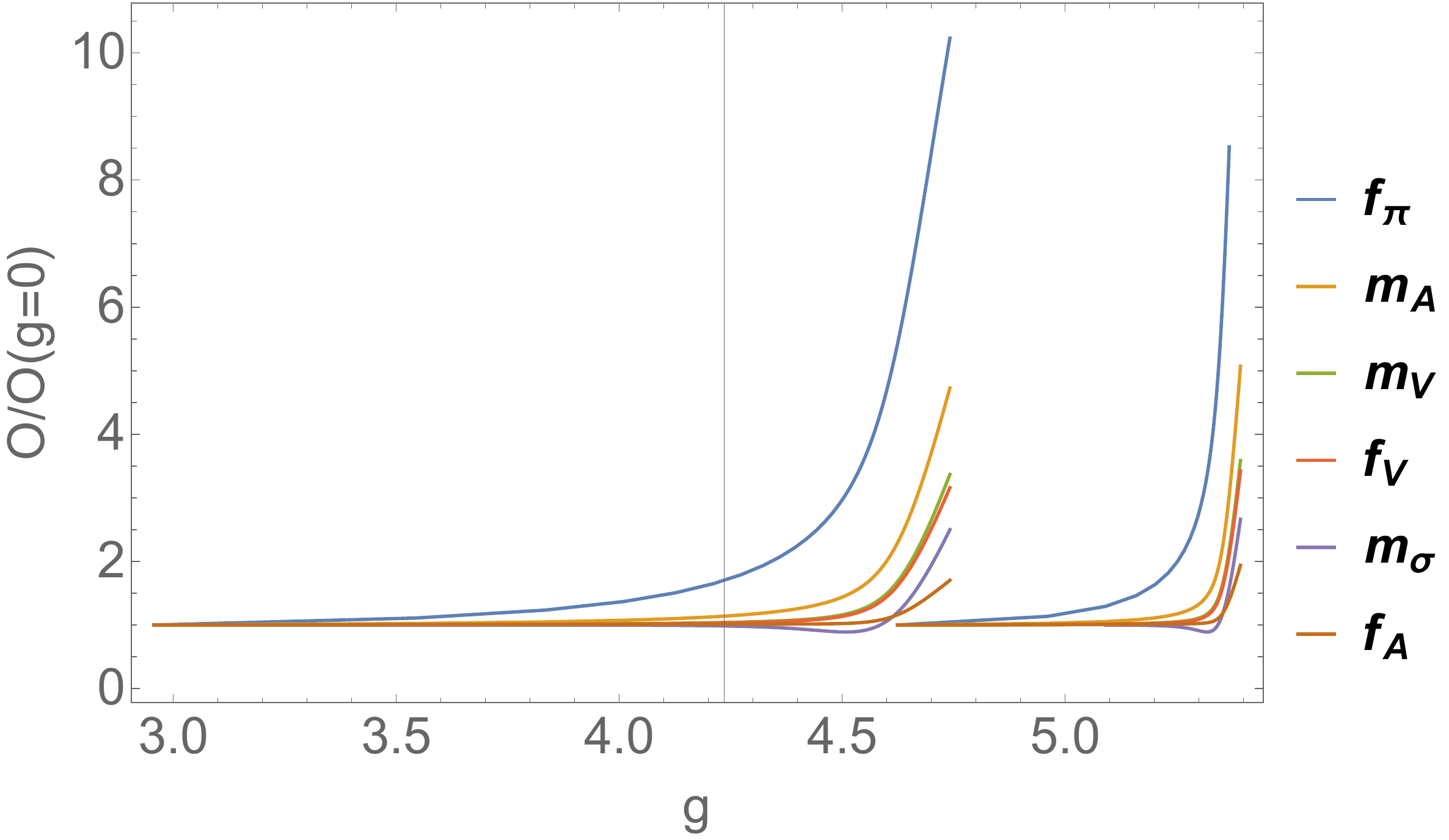}  
\caption{Plots showing the full set of observables against NJL coupling $g$ for $\Lambda=20$ and $50 \Lambda_{BF}$ . }
\label{lambda=20,50}
\end{figure}

\end{document}